\documentclass{svproc}
\usepackage{graphicx} 
\usepackage{amsmath}
\usepackage{pdfpages}
\usepackage[sorting=none]{biblatex}
\usepackage{mathptmx}

\title{Evaluating Ensemble Methods for News Recommender Systems}
\author{Alexander Gray, Noorhan Abbas}
\institute{University of Leeds, School of Computing, ODL MSc in AI, UK.}
\date{February 2023}

\begin{document}

\maketitle

\begin{abstract}

    News recommendation is crucial for facilitating individuals' access to articles, particularly amid the increasingly digital landscape of news consumption. Consequently, extensive research is dedicated to News Recommender Systems (NRS) with increasingly sophisticated algorithms. Despite this sustained scholarly inquiry, there exists a notable research gap regarding the potential synergy achievable by amalgamating these algorithms to yield superior outcomes. This paper endeavours to address this gap by demonstrating how ensemble methods can be used to combine many diverse state-of-the-art algorithms to achieve superior results on the Microsoft News dataset (MIND). Additionally, we identify scenarios where ensemble methods fail to improve results and offer explanations for this occurrence. Our findings demonstrate that a combination of NRS algorithms can outperform individual algorithms, provided that the base learners are sufficiently diverse, with improvements of up to 5\% observed for an ensemble consisting of a content-based BERT approach and the collaborative filtering LSTUR algorithm. Additionally, our results demonstrate the absence of any improvement when combining insufficiently distinct methods. These findings provide insight into successful approaches of ensemble methods in NRS and advocates for the development of better systems through appropriate ensemble solutions. 

\end{abstract}

\section{Introduction}

    In the UK, where certain news publications captivate 77\% of the online population \cite{Gazzette-Audience} and others produce almost 1500 stories a day \cite{Gazzette-output} the scale of online news cannot be overstated. But with this flood of readers and articles - how can the right story find the right person? 

    Recommender systems, which are algorithms designed to recommend items to users, have been deployed seemingly everywhere where recommendations are made - from movies to books to dating. With many of the largest corporations in the world including Amazon, Facebook, and Spotify all heavily featuring recommender systems, some studies claim that recommender systems are even responsible for up to a third of all traffic on the world's most popular sites \cite{rec_systems}.

    The types of recommendations that are made can be classified into three main categories based off their level of personalisation \cite{non-personalised}:

    \textbf{Non-personalized recommendations}: These recommendations are not customized to the user. They are often manually selected rather than algorithmically generated, such as editorial picks featured on a news front page or 'our favorites' highlighted on a retail site. Alternatively, they may be determined by key metrics like highest rated, most popular, or most recent items.
    
    \textbf{Ephemeral recommendations}: These recommendations are a form of personalised recommendation, but only tailored to what is in front of the user in the immediate present. For example, when making a purchase on Amazon recommendations may be made based on the items in the user's shopping basket. In the case of Amazon, these recommendations are often determined by either identifying the most similar items to those being purchased (using content-based algorithms) or by analysing which items are frequently bought together with the current selection in the basket.

    \textbf{Personalised recommendations}: Recommendations that are tailored to individual users or user groups and represent the primary focus of research in this field. These recommendations are generated based on the user's history, including articles read, items purchased, and other relevant interactions. Additionally, they may consider user demographics such as age or gender.

    The algorithmic approaches in news recommendation systems typically fall into three categories: content-based, collaborative filtering, or a hybrid approach. Content-based filtering approaches leverage features of the items to identify similar items to those the user already likes. In the context of news, this involves recommending similar articles to those the user has previously read. On the other hand, collaborative filtering techniques aim to find good recommendations by discovering users with similar preferences and suggesting items those users have found appealing. Hybrids, although not as common, are a mixture of the two approaches.

    An emerging domain within recommender systems is News Recommender Systems (NRS), which focuses on recommending news articles typically within news aggregations platform like Google News or Apple News, as well as on news publisher websites. NRS face various challenges specific to the domain, complicating the recommendation process. One such challenge, relevant for all recommender systems, but particularly for NRS is the cold-start problem. This issue refers to the challenge of making good recommendations for either new users or new items. It poses a difficulty as in both cases it means there is a lack of previous data indicating whether the items were well-received, either through implicit feedback (e.g. article clicks or item purchases) or explicit feedback (e.g. rating a movie or hotel with 5 stars). This presents a significant hurdle for news algorithms due to the sheer volume of new items, such as news articles, generated daily. 

    In this study, our focus will be on exploring ensemble approaches to personalised recommendations in the realm of news recommendation. This involves combining the outputs of individual distinct algorithms and assessing whether the combined performance can outperform the models individually. Specifically, we hypothesise that by combining well performing content-based and collaborative filtering algorithms that the result will be able to surpass both of their scores individually due of the distinct methodologies employed by these algorithms in calculating their recommendations.

    To our knowledge, this particular approach has not been explored within the news recommendation domain, nor has it been investigated in a recommendation domain that relies solely on implicit feedback.

\section{Related Work}

    \subsection{Datasets}
    
    Prior to the release of the Microsoft News Dataset (MIND) in 2020, progress in NRS was hindered by the absence of a comprehensive, extensive, and readily accessible dataset for evaluating novel algorithms \cite{Mind}. Among the few notable datasets available were Plista, comprising news data from 13 German news sites\cite{Plista}, Adressa, constructed by recording data from 10 weeks on Adresseavisen's news portal \cite{Adressa}, Globo which was news data provided by Globo.com \cite{Globo}, and Yahoo!, created by crawling news articles on Yahoo! News website \cite{Yahoo}. Table \ref{tab:datasets} provides details on these datasets including their published language, number of unique users, number of unique news articles, number of recorded clicks, and the news information available for each news article. 

    \begin{table}
    \centering
    \renewcommand{\arraystretch}{1.1} 
    \setlength{\tabcolsep}{5pt} 
    \begin{tabular}{|c|c|c|c|c|c|}
    \hline
    Dataset & Language & \# Users & \# News & \# Clicks & News Information \\
    \hline
    Plista & German & Unknown & 70,353 &  1,095,323 & title, body \\
    Adressa & Norwegian & 3,083,438 & 48,486 & 27,223,576 & title, body, category \\
    Globo & Portuguese & 314,000 & 46,000 & 3,000,000 & no original text, only word embeddings \\
    Yahoo! & English & Unknown & 14,180 & 34,022 & no original text, only word IDs \\
    \hline
    MIND & English & 1,000,000 & 161,013 & 24,155,470 & title, abstract, body, category \\
    \hline
    \end{tabular}
    \caption{Comparison of MIND dataset to existing public news datasets. \cite{Mind}}
    \label{tab:datasets}
    \end{table}

    The MIND dataset is compiled from user behaviour logs on the Microsoft News platform, recording data from 1 million distinct users and over 24 million clicks. Additionally, the dataset comprises of 161,000 unique English news articles each accompanied by titles, abstract, section, and subsection.   
    
    \subsection{News Recommendation Approaches}

    Traditionally, most methods relied on feature engineering to represent users and news. However, in recent years, there has been a notable shift towards collaborative-filtering neural network approaches \cite{CupMar}. This shift highlights the growing emphasis on data-driven methods for feature extraction and representing users and items allowing for the potential to capture interactions and patterns in greater detail. For example, CupMar \cite{CupMar} and Miner \cite{Miner} are two recent developments following this trend with both achieving state-of-the-art performance on the MIND dataset with AUC results of 0.6582 and 0.6961, respectively. As of September 2021, Miner ranked first on the MIND competition leaderboard, showcasing the robust performance of neural network approaches. The Miner utilizes a unique approach to user embeddings, employing a poly attention scheme to learn multiple interest vectors for each user, in contrast to the more conventional method of modelling user interest through a single user embedding.

    \subsection{Ensemble Recommender Systems}

    In the broader domain of recommender systems, there has been limited research on ensemble approaches. Gharahighehi et al. \cite{hypergraph} proposed an ensemble recommender system that used a unified hypergraph ranking framework, which modelled high-order relations to enhance recommendation accuracy. Additionally, in 2010, Jaher et al. \cite{jaher}. explored linearly combining various collaborative filtering algorithms on the Netflix Prize dataset, they were able to demonstrate the ensemble approach improved accuracy, outperforming all individual methods. 

    However, different from the previous methods, our approach aims to highlight the benefits of an ensemble approach of a variety of distinct methods relating to recommendation systems. Importantly, this study represents the first instance of such an ensemble approach being specifically applied within the news recommendation domain.

\section{Methodology}

    This study’s methodology encompasses three main subsections: Dataset, Base Learners, and Ensemble Approach. The Dataset subsection delineates the structure of the MIND dataset and provides a sample representation of its article data and user behaviour data. Base Learners elucidates the individual algorithms employed within the ensemble methods. Lastly, the Ensemble Approach details the rank aggregation methods utilized for combining recommendations generated by the individual algorithms into a cohesive ensemble ranked list. 

    \subsection{Dataset}

    The MIND dataset is structured as two main files. The first file operates at user interaction granularity and records the articles presented to the user, along with the previously clicked articles by that user. The second file provides information about each article contained within the dataset.

    Table \ref{tab:mind_article} presents a sample of the MIND article data. For each article that appears in a user's click history or is presented as a suggested article, the table records the article's designated section, subsection, the title, and the abstract of the article.
    
    \begin{table}
    \centering
    \renewcommand{\arraystretch}{1.1} 
    \setlength{\tabcolsep}{5pt} 
    \begin{tabular}{|c|c|c|c|c|}
    \hline
    id & section & subsection & title & abstract \\
    \hline
    N55528 & lifestyle & lifestyleroyals & The Brands Queen Eliz... &  Shop the notebook...\\
    N19639 & health & weightloss & 50 Worst Habits For Belly Fat & These seemingly harml... \\
    N61837 & news & newsworld & The Cost of Trump's Aid Freeze i... & Lt. Ivan Molchanets...\\
    \hline
    \end{tabular}
    \caption{Snippet of MIND's news article data}
    \label{tab:mind_article}
    \end{table}

    Table \ref{tab:mind_behaviour} displays a sample of the MIND behaviours data. It logs each user session at a specific time, providing the user's click history leading up to that point and the impression article IDs shown to the user. The IDs are appended with "-0" to signify that the article was not clicked and "-1" to indicate that it was clicked.
    
    \begin{table}
    \centering
    \renewcommand{\arraystretch}{1.1} 
    \setlength{\tabcolsep}{5pt} 
    \begin{tabular}{|c|c|c|c|c|}
    \hline
    id & user id & time & click history & impressions \\
    \hline
    1 & U13740 & 11/11/2019 9:05:58 AM & N55189 N42782 N34694... &  N55689-1 N35729-0\\
    2 & U91836 & 11/12/2019 6:11:30 PM & N31739 N6072 N63045... & N20678-0 N39317-0... \\
    3 & U73700 & 11/14/2019 7:01:48 AM & N10078 N56514 & N39985-0\\
    \hline
    \end{tabular}
    \caption{Snippet of MIND behaviours data}
    \label{tab:mind_behaviour}
    \end{table}

    Since its publication in 2020, the MIND dataset has garnered considerable attention, receiving at the time of writing over 400 citations \cite{aclanthology2020}, with the majority of novel NRS methods measuring their results using the dataset, including Miner\cite{Miner}, LSTUR \cite{lstur}, NPA \cite{npa}, among others.
    
    \subsection{Base Learners}

        Described below are the Base learners evaluated within the ensemble methods. Base learners represent individual algorithms that are combined to form the ensemble approach.

        \subsubsection{Neural News Recommendation with Long-term and Short-term User Representations (LSTUR)}

        A neural network based recommendation model designed to capture both long-term and short-term user interests for news recommendation \cite{lstur}. This model employs a blend of recurrent and convolutional neural networks to encode user behaviour sequences and article features, respectively, enabling the generation of personalised recommendations.
    
        \subsubsection{Neural News Personalisation with Auxiliary Information (NPA)}

        A recommendation model which integrates user demographics, user-item interactions, and content features to improve recommendation performance \cite{npa}. The model utilises a multi-task learning framework to jointly optimize recommendation accuracy and auxiliary prediction tasks.
    
        \subsubsection{Neural News Recommendation with Multi-Head Self-Attention (NRMS)}

        A news recommender approach that combines a neural network model with attention mechanisms in an attempt to capture user preference to deliver recommendations \cite{nrms}. The key feature is the use of multi-head self-attention that allows the model to simultaneously attend to different parts of the news articles and effectively capture their semantic representations. This enables NRMS to grasp more nuanced user-article relations for accurate, personalised recommendations.

        \subsubsection{Term Frequency-Inverse Document Frequency (TF-IDF)}

        A content-based approach, term frequency-inverse document frequency can measure the similarity between a concatenation of the headline and the abstract of each article. A simple version of this approach would be to compare if the same words occur in the two entries, however TF-IDF places greater value to words that rarely appear in the total set of items and lower value to words that appear in more of them. This results in the word "Queen" equating to much higher similarity values compared to the word "The" between two articles when matched.

        TF-IDF can be calculated as shown in \ref{tfidf} which is the dot product between term-frequency and inverse document frequency (IDF). Term frequency \ref{tf} is calculated through finding the how often the term appears in the document relative to the total number of terms in the document. IDF can be calculated, as shown in \ref{idf}, which measures the significance of a term by inversely scaling its frequency across all documents, calculating for an individual term t as the logarithmic ratio between total number of documents (N) and the number of documents containing term t (df(t)).

        \begin{equation}
        \text{tf-idf(t,d,D)} = \text{tf(t,d)} \cdot \text{idf(t,D)}
        \label{tfidf}
        \end{equation}

        \begin{equation}
        \text{Term Frequency} = \frac{\text{Number of times the term appears in the document}}{\text{Total number of terms in the document}}
        \label{tf}
        \end{equation}

        \begin{equation}
        \text{IDF}(t) = \log\left(\frac{N}{\text{df}(t)}\right)
        \label{idf}
        \end{equation}

        \subsubsection{Bidirectional Encoder Representations from Transformers (BERT)}

        Another content-based approach, BERT uses encoders to convey the meaning of the words in an n-dimensional space \cite{bert}. This approach offers an advantage over traditional word-based methods because semantically similar words, such as "King" and "Queen," will appear more similar in this vector space than they would through a simple string comparison.

    \subsection{Ensemble approach}

        The individual algorithms produce their recommendations in the form of an ordered list such as [5,1,2,3,4] where each number represents an item and their order signifies the recommendation priority, with 1 being the most recommended item and n being the least recommended. Therefore, assembling an ensemble of these rankings presents a rank aggregation challenge. We utilize a linear combination approach, where the rankings generated by individual algorithms are combined linearly by aggregating their scores to form a weighted list. This weighted list is subsequently sorted to yield a final ranked list as the final output.

    \subsection{Performance metrics}

    This paper employs a trio of metrics to evaluate the performance of both the base learners and ensemble algorithms. Since the output of the algorithms is in the form a ranked list, the metrics employed are decision support metrics rather than prediction accuracy metrics. This means that they assess the algorithms based on ability to recommend relevant items rather than focusing solely on the correctness of individual recommendations.

        \subsubsection{Area under the ROC Curve (AUC)}

        The AUC curve measures how well a user will like recommended items or not. It is computed through plotting the False Positive Rate (FPR) against the True Positive Rate (TPR) and subsequently calculating the area under the resulting curve. The value will fall between the range of 0 and 1, where 0.5 indicates random guessing. The equations for calculating TPR and FPR are shown in equations \ref{tpr} and \ref{fpr}, respectively.

        \begin{equation}
        \text{True Positive Rate} = \frac{\text{True Positives}}{\text{True Positives} + \text{False Negatives}}
        \label{tpr}
        \end{equation}

        \begin{equation}
        \text{False Positive Rate} = \frac{\text{False Positives}}{\text{False Positives} + \text{True Negatives}}
        \label{fpr}
        \end{equation}

        \begin{figure}[h]
        \includegraphics[width=8cm]{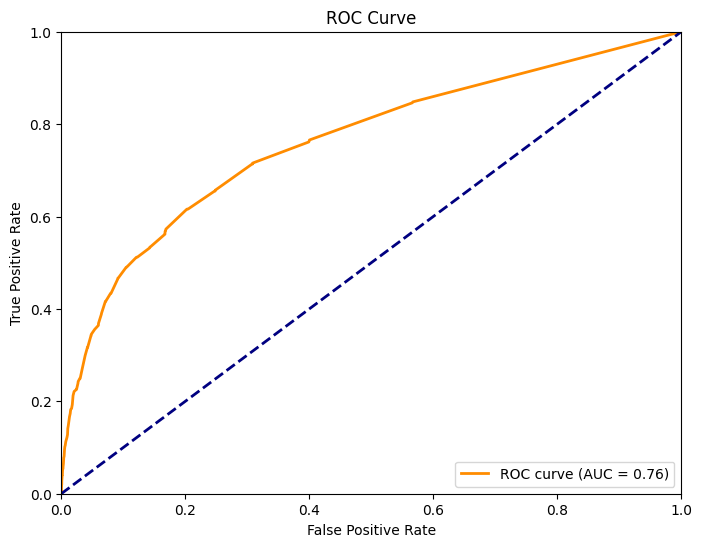}
        \centering
        \caption{Example of an ROC curve}
        \end{figure}
    
        \subsubsection{Mean Reciprocal Rank (MRR)}

        MRR is specific to search or recommender problems and it calculates the average of the reciprocal of the rank of the first relevant item retrieved for a set of queries and the formula is shown in equation \ref{mrr}. In the formula, \( |U| \) represents the total number of queries, rank denotes the position of the first relevant item in the ranked list for query \( u \), and the summation aggregates these values across all queries. This metric assesses how effectively a search or recommendation system places relevant items at the top of the list.
        
        \begin{equation}
        \text{MRR} = \frac{1}{|U|} \sum_{u=1}^{|U|} \frac{1}{\text{rank}_u}
        \label{mrr}
        \end{equation}
    
        \subsubsection{Normalized discounted cumulative gain (NDCG)}

        Normalized discounted cumulative gain is a widely used metrics for evaluating search or recommender systems.  It assesses the quality of ranked list of items, considering the relevance and position of each item in the list. For evaluation of the models we consider two forms of NDCG, the first which considers the "top 5" positions on the produced ranked list (NDCG@5) and the second which considers the "top 10" (NDCG@10). The formula is shown in equation \ref{NDCG} where DCG\textsubscript{u} represents the discounted cumulative gain for a user - a measure of the relevance of the items in each position in a ranked list. IDCG\textsubscript{u} is the ideal DCG\textsubscript{u} which is the greatest value that is achievable for that user if all items were ranked in the perfect order, and $\vert$U$\vert$ is the total number of users in the evaluation dataset.

        \begin{equation}
        \text{NDCG}@k = \frac{1}{|U|} \sum_{u=1}^{|U|} \frac{DCG_u}{IDCG_u}
        \label{NDCG}
        \end{equation}
        
    \subsection{Hyperparameters}
    
    For all collaborative filtering base learners in the study, parameter tuning was not required as optimal configurations had been predefined and evaluated using the same MIND dataset. For the BERT transformer, a comprehensive selection of hyperparameters was undertaken, encompassing crucial variables such as max\_length, truncation, batch\_size, and learning\_rate. Through a grid search methodology, their performance was assessed to ascertain the optimal configuration, thereby determining the most effective combination of parameters.

    \subsection{Implementation}

    The proposed methods were implemented using Python 3.8. The implementation made use of several Python libraries, including Recommenders for the neural network approaches, Scikit-learn for various data preprocessing tasks throughout, and Pandas and NumPy for data manipulation. Git versioning control was employed throughout the coding process, enabling all changes to be tracked through a git history with detailed commit messages.

\section{Results}

    Initially, we studied the performance of an ensemble approach employing multiple collaborative filtering methods, namely NPA, LSTUR, and NRMS. The results are shown in table \ref{tab:Collaborative-filtering ensemble results}.

    \begin{table}
    \centering
    \renewcommand{\arraystretch}{1.1} 
    \setlength{\tabcolsep}{5pt} 
    \begin{tabular}{|c|c|c|c|c|}
    \hline
    Model & AUC & MRR & nDCG@5 & nDCG@10 \\
    \hline
    NPA & 0.58 & 0.25 & 0.27 & 0.34 \\
    LSTUR & 0.60 & 0.26 &  0.29 & 0.35 \\
    NRMS & 0.58 & 0.24 & 0.26 & 0.33 \\
    Combined & 0.60 & 0.25 & 0.27 & 0.34 \\
    \hline
    \end{tabular}
    \caption{Collaborative-filtering ensemble results}
    \label{tab:Collaborative-filtering ensemble results}
    \end{table}

    These results show that the ensemble of these individual methods was not able to surpass the performance of the best model individually. This supports the statement that for ensemble methods to be beneficial there has to be a significant degree in diversity of the base learners, which in this context demonstrates that these methods approach are too similar - all being neural network approaches, albeit with still diverse approaches apart from that, collaborative filtering methods.
    
    Next, we then looked specifically at the performance of an ensemble of the two content-based methods - TF-IDF and the BERT approach - expecting to have similar results to the previous trial as both methods are attempting to rank the suggested articles in order of the similarity to previously read articles with the BERT approach doing it semantically and TF-IDF doing it through comparing the frequency of the words.

    \begin{table}
    \centering
    \renewcommand{\arraystretch}{1.1} 
    \setlength{\tabcolsep}{5pt} 
    \begin{tabular}{|c|c|c|c|c|}
    \hline
    
         Model & AUC & MRR & nDCG@5 & nDCG@10 \\
         \hline
         TF-IDF & 0.58 & 0.27 & 0.29 & 0.35 \\
         BERT & 0.60 & 0.27 &  0.29 & 0.35 \\
         Combined & 0.60 & 0.28 & 0.30 & 0.37 \\
    \hline
    \end{tabular}
    \caption{Content-based ensemble results}
    \label{tab:Content-based ensemble results}

    \end{table}

    The findings shown in table \ref{tab:Content-based ensemble results} further the assertion made in the preceding results and in the initial statement that for ensembles to yield beneficial results, there must exist a high degree of diversity between the constituent models. Such diversity is likely to require an ensemble of a collaborative filtering and content-based models. While these results exhibit a slight enhancement in metrics such as MRR, nDCG@5, and nDCG@10, the improvements are not substantial enough to warrant the adoption of an ensemble approach.

    The final test outcomes are depicted in table \ref{tab:Final testing}. In this analysis, we examined the result of combining the best-performing collaborative filtering method with the best-performing content-based method. These results demonstrate that the combination of the best performing collaborative filtering method and the best performing content-based method were able to vastly surpass both of the individual methods. These results support the initial hypotheses that an ensemble of recommendation system algorithms could yield results superior to the models individually, given that the models were distinct enough.

\begin{table}
    \centering
    \setlength{\tabcolsep}{5pt}
    \begin{tabular}{|c|c|c|c|c|}
        \hline
         Model & AUC & MRR & nDCG@5 & nDCG@10 \\
         \hline
         Bert & 0.60 & 0.27 &  0.29 & 0.35 \\
         LSTUR & 0.60 & 0.26 &  0.29 & 0.35 \\
         Combined & 0.63 & 0.28 &  0.31 & 0.37 \\
         \hline
         
    \end{tabular}
    \caption{Final testing results}
    \label{tab:Final testing}
\end{table}

\section{Conclusion}

This study demonstrates how ensemble methods of diverse state-of-the-art recommender system algorithms can results in improved performance of those individually. The results demonstrate that by an appropriate choice of base learners that the ensemble yields superior results seemingly proportionally to how distinct the base learners are. This is supported by the results which showed an ensemble of numerous neural network collaborative filtering algorithms could not yield results superior to the best performing algorithm, and similar results for the content-based approach where the results showed a very marginal increase in some metrics for two fairly similar approaches. Then when running an ensemble of the best performing collaborative filtering and content-based approaches this resulted in a \% improvements in AUC and significant improvements in all metrics. Not only does this demonstrate how optimal ensemble algorithms can be deployed for News Recommendation Systems but also that existing state-of-the-art collaborative filtering algorithms could be even further improved through an ensemble with leading content-based approaches.

\subsection{Ethical Issues}

One ethical issue with this approach is the potential lack of focus on diversity or serendipity. This could exacerbate the echo-chamber effect by consistently recommending users content that aligns with their existing opinions and perspectives, without providing exposure to contrasting viewpoints.

Furthermore, since collaborative filtering methods operate by identifying similar users based on context such as user's demographics, without adequate safeguards in place, this could result in the algorithm suggesting stereotypical items to marginalized groups, thereby reinforcing existing stereotypes.

Moreover, within the context of news aggregation, a fairness concern arises for news publishers. Platforms such as Google News and Google Discover rely on recommender system and serve as a major traffic source for news publishers. These algorithms essentially determine which news channels receive more traffic and, consequently, more revenue.  This issue could be exaggerated if the algorithm exhibits biases toward individual websites or if the algorithm could be manipulated by news providers.

\subsection{Future Work}

The assessment of results relied on the MIND dataset, categorized as "offline" evaluation which some research indicates that it may not directly correlate with online settings. \cite{offline-online-eval} Hence, albeit beyond the scope of this study, there exists the prospect of validating these findings in an online environment—by assessing the amalgamation of cutting-edge collaborative filtering and content-based algorithms on a live news platform in real-time.

Additionally, given the study's findings revealed that greater diversity in the base learners led to better results in the ensemble algorithm, there is an opportunity to reinforce this observation with further testing of a variety of state-of-the-art algorithms with a number of unique approaches.

Finally, as we also only experimented with linearly combining the base learners, there is also room to experiment with a greater variety of ensemble methods such as non-linear combination techniques like stacking or blending, hierarchical ensemble structures, and adaptive weighting schemes, or more advanced ensemble methods such as hypergraphs. By exploring a broader spectrum of ensemble methodologies, greater approaches tailored to thee specific characteristics of the datasets and base learners can be found to enhance overall predictive performance.

\printbibliography
\end{document}